\documentclass[%
	aip,
	amsmath,amssymb,
	reprint,
]{revtex4-2}

\usepackage{graphicx}
\usepackage{dcolumn}
\usepackage{bm}
\usepackage[utf8]{inputenc}
\usepackage[T1]{fontenc}
\usepackage{mathptmx}
\usepackage{xcolor} 
\usepackage{float}

\graphicspath{{gfx/}}

\renewcommand{\i}{\text{i}}
\renewcommand{\d}{\text{d}}
\newcommand{\E}{\mathcal{E}}
\newcommand{\br}[1]{\langle #1 \vert}
\newcommand{\ke}[1]{\vert #1 \rangle}

\newcommand{\defeq}{:=}
\newcommand{\D}{\mathfrak{D}}

\begin{document}

	\title{Simulating Photodissociation Reactions in Bad Cavities with the Lindblad Equation}
	
	\author{Eric Davidsson}
	\email{eric.davidsson@fysik.su.se}
	\author{Markus Kowalewski}%
	\email{markus.kowalewski@fysik.su.se}
	\affiliation{Department of Physics, Stockholm University, Albanova University Center, SE-106 91 Stockholm, Sweden}
	
	\date{\today}

	\begin{abstract}
		Optical cavities, e.g.\ as used in organic polariton experiments, often employ low finesse mirrors or plasmonic
		structures. 
		The photon lifetime in these setups is comparable
		to the timescale of the nuclear dynamics governing the photochemistry. 
		This highlights the need for including the effect of dissipation in the molecular simulations.
		In this study, we perform wave packet dynamics with the Lindblad master equation,
		to study the effect of a finite photon lifetime on the dissociation of the MgH$^+$ molecule model system.
		Photon lifetimes of several different orders of magnitude are considered
		to encompass an ample range of effects inherent to lossy cavities.
	\end{abstract}
	
	\maketitle
	
	\section{Introduction}
	Motivated by recent experimental advancements,\cite{yuen-zhou-etal-2018,schwartz-etal-2011,thomas-etal-2019}
	state-of-the-art computational tools are presently employed to consider
	the interaction between quantized electromagnetic radiation,
	and a diverse range of matter systems,
	in models reminiscent of the early Jaynes-Cummings model.\cite{jaynes-cummings-1963}
	Examples of such matter systems include
	quantum dots,\cite{jia-2018}
	excitons,\cite{byrnes-2014}
	ensemble systems,\cite{hoi-ling-etal-2017,vendrell-2018,davidsson-kowalewski-2020}
	Bose-Einstein condensates,\cite{brennecke-2007}
	superconductors,\cite{baranov-etal-2018}
	and molecules of different sizes and complexity.\cite{flick-narang-2018}
	
	Central to these studies,
	be it experimental or theoretical,
	is the macro- or nanoscopic material structure
	giving rise to electromagnetic field-modes in a confined space,
	henceforth called a cavity structure,
	though they are not always reminiscent of the widely used Fabry-Pérot cavity.
	The material and geometry of these cavity structures determine the characteristics of the light-matter interaction,
	and precise tailoring enables a unique possibility to probe or control phenomena 
	that take place down on the scale of single particles.
	Examples of such phenomena include,
	chemical reactions,\cite{yuen-zhou-etal-2018}
	photo-dissociation,\cite{davidsson-kowalewski-2020}
	electron transport,\cite{herrera-spano-2016,cortes-2017}
	isomerization,\cite{mandal-huo-2020}
	and heat-transfer between molecules.\cite{ashrafi-etal-2020}
		
	In theoretical works, such as this,
	the properties of the cavity structure
	are typically translated into parametric values,
	such as the mode frequency and the vacuum electric field strength,\cite{flick-etal-2017-pnas,kowalewski-mukamel-2017}
	entering the model of the system.
	The influence of these parameters are well studied and generally well understood.
	
	However, not all parameters describing the cavity structure have gained equal attention.
	Effects arising from a finite lifetime for field excitations,
	a.k.a. cavity Q-factor or photon decay-rate,
	have only recently gained attention as the main focus in a few ground-breaking studies,
	\cite{ulusoy-vendrell-2020,antoniou-etal-2020,felicetti-etal-2020}
	and dissipative effects in general, have been described as a challenge for computational methods.\cite{yuen-zhou-etal-2018}
	
	Theoretical studies often assume an infinite lifetime,\cite{mandal-huo-2020}
	which can be appropriate
	but introduces inaccuracies when used to model conditions 
	where finite photon lifetimes are de facto non-negligible.
	For instance, this is typically the case for plasmonic nano-cavity structures,
	\cite{hugall-etal-2018,baranov-etal-2018}
	such as ultra-thin metallic gaps\cite{baumberg-etal-2019} or nano-gap antennas,\cite{li-etal-2016,climent-etal-2019}
	where the desire to concentrate a strong electromagnetic field in a small volume 
	inevitably leads to short-lived excitations of the field.\cite{yuen-zhou-etal-2018}
	
	Another previously employed strategy is to incorporate a Hamiltonian coupling between affected states and a reservoir of simple systems.\cite{breuer-petruccione-2010}
	However, this allows the energy to return from the reservoir and affect the evolution of the system,
	which therefore makes an approximative model for electromagnetic energy being lost into the environment.
	Additionally, photon decay processes inevitably introduce decoherence into the system,
	which makes an approach with a pure state wave function insufficient.\cite{manzano-2020}
	Instead, a density matrix formalism is required for time evolution of these statistically mixed states,
	and the Schr\"odinger equation is generalized to some variety of a master equation.
	Here we use the Markovian Lindblad master equation,
	where the decay-rate of field excitations will enter through the cavity decay rate $\kappa$.\cite{manzano-2020,Haroche2006}
	\begin{equation}
		\label{eq:lindblad}
		\partial_t \hat \rho
		=
		-\frac{\i}{\hbar} \big[ \hat H \! , \hat \rho \big]
		+
		\sum_{n} \kappa_{n} \Big(
			\hat L_{n}\hat \rho \hat L^\dagger_{n}
			- \frac{1}{2}
			\big[ \hat L_{n}^\dagger \hat L_{n}, \hat \rho \big]_{+}
		\Big)
	\end{equation}
	However, using the density matrix, $\hat \rho$,
	increases the computational cost considerably compared to wave function based approach.
	
	Recent publications\cite{ulusoy-vendrell-2020,fregoni-etal-2018,fregoni-etal-2020,antoniou-etal-2020,felicetti-etal-2020} 
	with cavity decay do not deal directly with the Lindblad equation formalism;
	instead, the underlying formalism is a non-Hermitian evolution of a pure state,
	that is shown to be appropriate both
	for relaxation dynamics,\cite{ulusoy-vendrell-2020,felicetti-etal-2020} and isomerization.\cite{antoniou-etal-2020}
	However, for general applications,
	the method omits some phenomenological processes,
	such as decoherence and time evolution of statistically mixed states,
	and the impact of this exclusion depends on the type of system and observable under investigation.
	
	In this paper, we investigate the molecular dissociation of the MgH$^+$ molecule
	after excitation to an unbound electronic state,
	and in the presence of a lossy cavity structure.	
	Here the photon decay will populate intermediate states that also contribute
	to the dissociation. Thus a wave function based approach is not suitable for this type of problem.
	(See discussion at the end of section \ref{sec:system-model}).
	Instead, the Lindblad master equation is used to capture the behavior of the system 
	and deliver accurate results.
	We investigate how the photon lifetime and cavity vacuum field strength 
	affects the photostability of the MgH$^+$ molecule,
	and the photochemical reaction mechanisms of the coupled light-matter system are analyzed.
	
	\section{System and model}
	\label{sec:system-model}	
\begin{figure}
		\centering
		\includegraphics[width=\columnwidth]{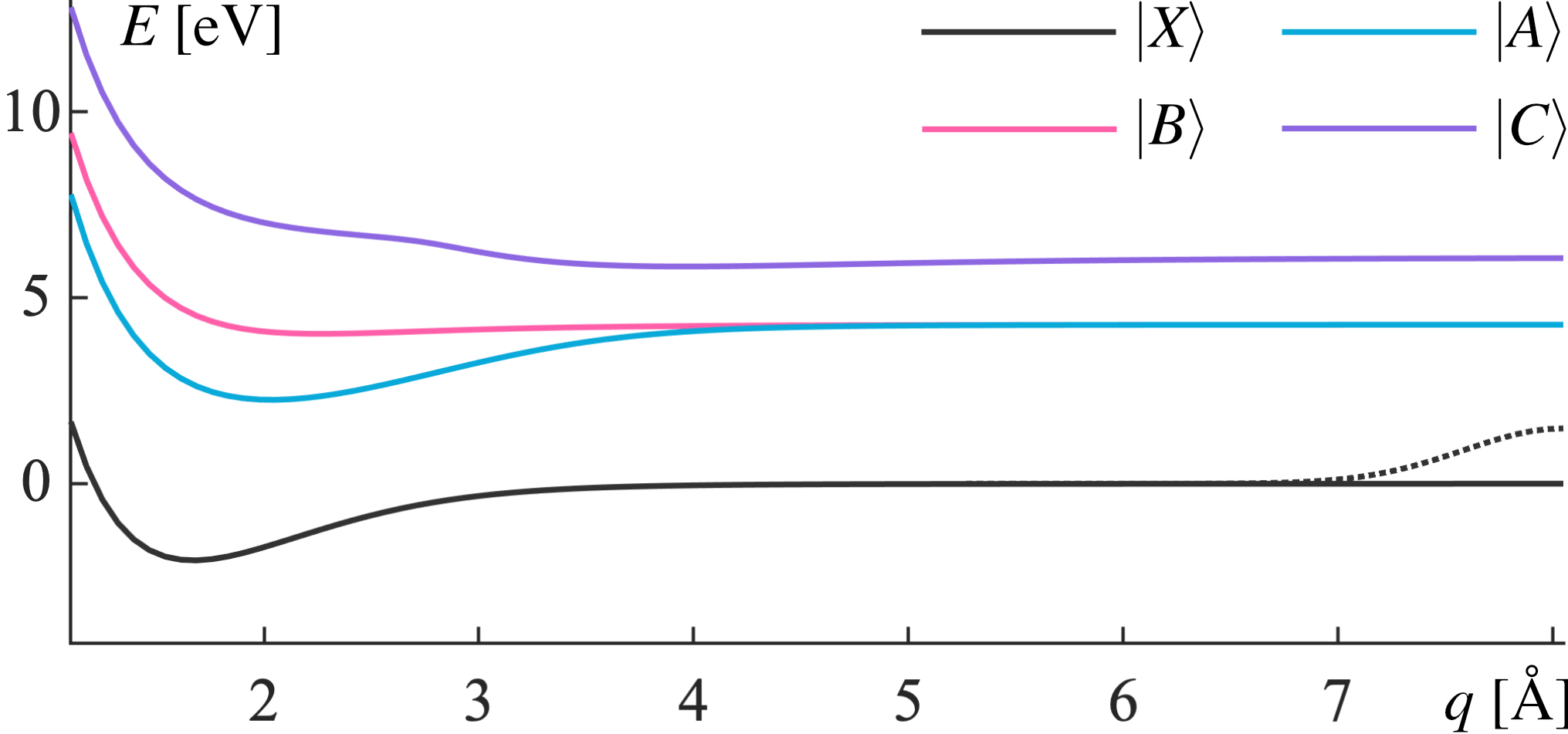}
		\caption{The four lowest electronic states in MgH$^+$.
		Each potential energy surface is implemented with an absorbing potential,
		whose Gaussian shape and relative size are shown with the dotted line on the black curve.
		\label{fig:pes}}
	\end{figure}

	The Hamiltonian in Eq.\ \eqref{eq:lindblad} is a molecular Jaynes-Cummings type Hamiltonian
	\cite{davidsson-kowalewski-2020}
	modeling vibrationally and electronically excited states in the presence of a lossy cavity mode.
	It is comprised of $\hat H_m$ for the molecule,
	$\hat H_c$ for the cavity mode,
	and the light-matter interaction $\hat H_{cm}$:
	\begin{equation}
		\label{eq:hamiltonian}
		\hat H
		=
		\hat H_m + \hat H_c + \hat H_{cm}
	\end{equation}
	We assume the cavity Born-Oppenheimer approximation,\cite{flick-etal-2017}
	the rotating wave approximation, and the dipole approximation,
	for a spatially fixed molecule.
	Excited electronic and vibrational states of the molecule are described 
	on one-dimensional potential energy surfaces.
	The four lowest electronically excited states of MgH$^+$ are considered (see Fig.\ \ref{fig:pes}).
	Wave-packets approaching the dissociation limit are absorbed by an imaginary potential.
	The cavity mode is modeled as 
	a single mode with a photon energy of 4.3\,eV (285\,nm).
	Comprehensive details about the Hamiltonian are found in the appendix,
	section \ref{sec:hamiltonian}.
	
	The states of the the combined molecule-cavity system are
	expressed as a product state,
	$\ke{n,M} = \ke{n} \otimes \ke{M}$,
	where $\ke{n} \in \{\ke{0}, \ke{1}, \ke{2}, \cdots\}$ are the Fock states
	of the cavity mode and $\ke{M} \in \{\ke{X}$, $\ke{A}$, $\ke{B}$, $\ke{C}\}$
	are the electronic states of MgH$^+$.
	All product states that have distinctly higher energy than the initial state,
	$\ke{0,C}$,	will never be populated under the rotating wave approximation,
	and are removed from the description.
	The full Hamiltonian, in Eq.\ \eqref{eq:hamiltonian}, is then expressed in the basis of eight states,
	$\{\ke{0,X}$, $\ke{1,X}$, $\ke{0,A}$, $\ke{0,B}$,
	$\ke{2,X}$, $\ke{1,A}$, $\ke{1,B}$, $\ke{0,C}\}$,
	covering a an energy range of $\sim10$\,eV.
	
	\begin{figure}
		\centering
		\includegraphics[width=\columnwidth]{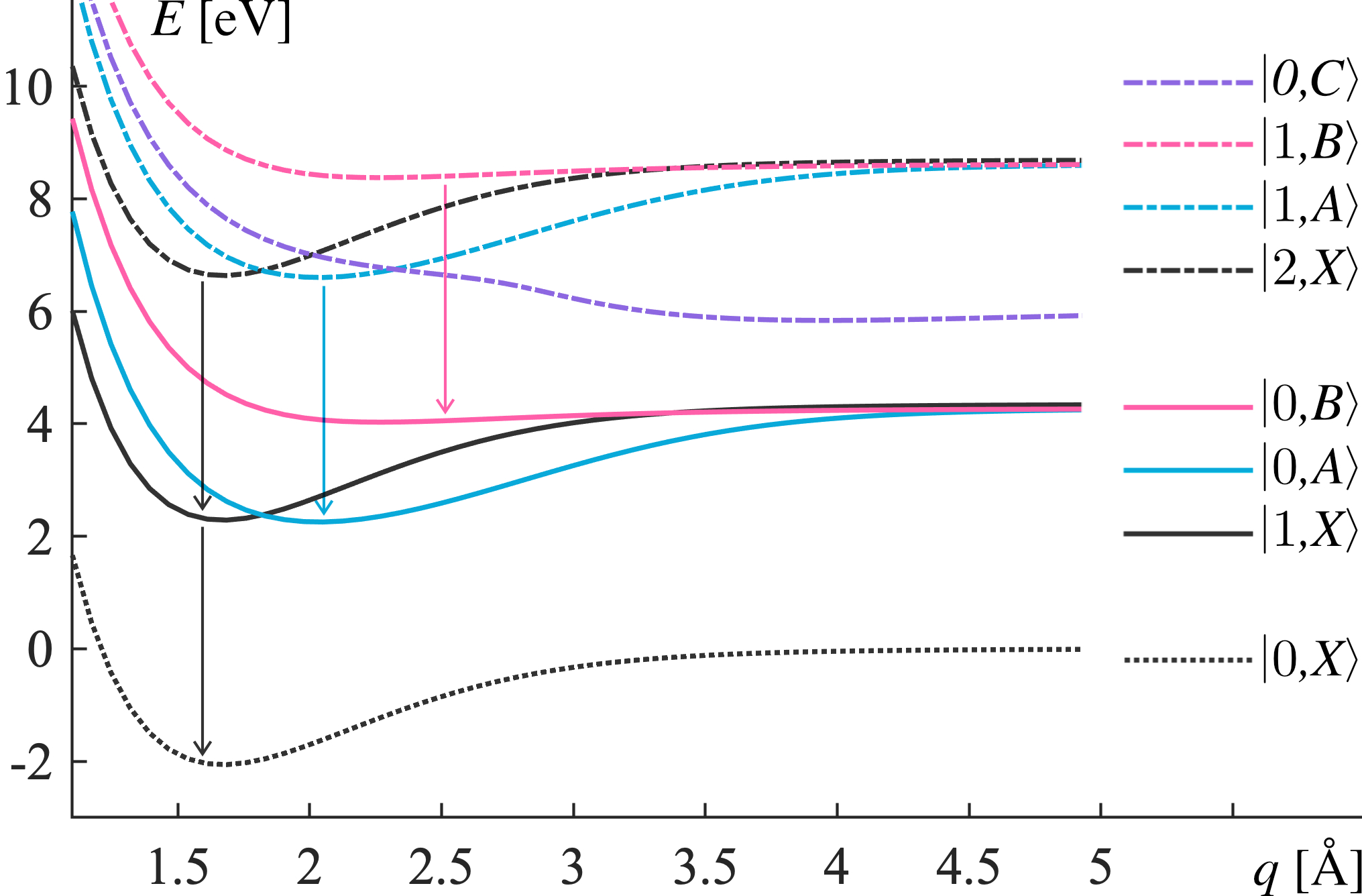}
		\caption{
		Potential energy surfaces associated with the basis-states for electronic and cavity degrees of freedom.
		The initial state is the pure molecular excitation, $\ke{0,C}$.
		Arrows indicate which states are coupled by Lindblad decay operators.
		States shown as dashed curves are referred to as the double excitation subspace.
		States shown as solid curves are referred to as the single excitation subspace.
		The state shown as a dotted curve is referred to as the ground-state subspace.
		The two lower subspaces (ground state and single excitation) 
		are occasionally considered as a group
		and referred to as the decohering subspaces.
		\label{fig:basis}}
	\end{figure}
	
	The product states are grouped into three subspaces:
	the ground state subspace,
	the single excitation subspace,
	and the double excitation subspace.
	The corresponding potential energy curves are shown in Fig.\ \ref{fig:basis}.	
	This partition will later be used for analyzing the dynamics of the system.
	Under the rotating wave approximation states from different excitation subspaces
	are not coupled, which simplifies the Hamiltonian \cite{kowalewski-etal-2016}.
	Only states within the same excitation subspaces are coupled and exhibit curve crossings
	as shown in Fig.\ \ref{fig:basis}.		
	
	The photon decay, as modeled by the Markovian Lindblad master equation,
	assumes that the photon is lost irrevocably from the system \cite{Haroche2006} (either into free space
	or by absorption at the surrounding bulk material).
	
	In the product basis,
	these unidirectional interactions are identified by, wherever possible,
	decreasing the photon number by one.
	This yields four transitions:
	$\ke{1,B} \!\rightarrow\! \ke{0,B}$,
	$\ke{1,A} \!\rightarrow\! \ke{0,A}$,
	$\ke{2,X} \!\rightarrow\! \ke{1,X}$, 
	and $\ke{1,X} \!\rightarrow\! \ke{0,X}$,
	which are indicated by vertical arrows in Fig.\ \ref{fig:basis}.
	Each of these unidirectional interactions forms a non-Hermitian Lindblad jump operator.
	The decay rates $\{\kappa_n\}$ from the general case in Eq.\ \eqref{eq:lindblad} 
	are here the same value,
	$\kappa$, for all photons in the mode,
	and $\kappa$ is determined for any particular cavity structure.
	Since the state $\ke{2,X}$ has two photons,
	the associated annihilation operator will introduce a factor of 2 in the Lindblad operator,
	effectively increasing the decay-rate proportional to the energy increase in the field.
	\begin{equation}
		\left\{
			\begin{alignedat}{2}
				&L_1 = &&\ke{0,B} \! \br{1,B} \\
				&L_2 = &&\ke{0,A} \! \br{1,A} \\
				&L_3 = 2&&\ke{1,X} \! \br{2,X} \\
				&L_4 = &&\ke{0,X} \! \br{1,X}
			\end{alignedat}
		\right.
	\end{equation}
	This model includes no Lindblad de-phasing operators,
	such as effects derived from weak interactions between the molecule and its environment,
	which are assumed to be negligible on the timescale for molecular dissociation of MgH$^+$.
	
	The initial state for the simulation, $\hat \rho_0$,
	is formed from the pure state $\ke{0,C,\psi(q)}$,
	where the component $\psi(q)$ is the nuclear vibrational wave function,
	created by weighting the ground-state wave-function of $\ke{X}$
	with the transition dipole moment between ${\ke{X} \! \leftrightarrow \! \ke{C}}$, which
	corresponds to a vertical excitation.
	\begin{equation}
		\hat \rho_0 = \ke{0,C,\psi(q)}\br{0,C,\psi(q)}
	\end{equation}
	This method provides a consistent initial condition of a fully exited molecule
	without the introduction of additional parameters.
	With $\hat H$, $\{\hat L_{n}\}$, $\kappa$, and $\hat \rho_0$,
	the model is fully defined,
	and the time evolution of Eq.\ \eqref{eq:lindblad} can be simulated
	as described in the Method section \ref{sec:method}.
	
	As a general remark:
	To minimize computational cost,
	the Lindblad master equation can be reduced 
	to a non-Hermitian Schrödinger equation employing an absorbing potential,
	given three conditions:
	The Hamiltonian does not couple any state in the subspace to other states that are not in the subspace,
	the initial state projected onto that subspace is a pure state,
	and the subspace is not the recipient of decaying states.
	This motivates the method in previous studies,
	\cite{ulusoy-vendrell-2020,antoniou-etal-2020,felicetti-etal-2020}
	but in this study, these conditions only apply to the doubly excited subspace.
	(See Fig.\ \ref{fig:basis} for definitions of subspaces.)
	The main issue is that a straightforward reduction to a Schrödinger equation
	only accounts for removal of population,
	and not the re-population that happens in the single excitation subspace,
	and ground-state subspace.
	The re-population is necessary for this study since a dominant contribution to the observable occur after it.
	Additionally, when the decay from the Lindblad operators is transferring population between subspaces,
	the state is decohering
	and re-populating the receiving subspaces as a statistical mixture of states.
	These mixed states display a suppressed interference when evolving on the potential energy surfaces,
	an effect that is further discussed in relation to our data in section~\ref{sec:results-discussion}.
	For these reasons, time evolution of a non-Hermitian Schrödinger equation,
	which only accommodates pure states,
	is not well suited to our system.
	But instead, the Lindblad equation will describe these influential effects.

	\section{Methods}
	\label{sec:method}
	
	The potential energy curves, shown in Fig.\ \ref{fig:pes},
	are calculated with the program package  Molpro\cite{molpro-2006} 
	at the CASSCF(12/10)/MRCI/ROOS level of theory.
	\cite{widmark-1990,widmark-1991}	
	The time evolution of the Lindblad equation is done numerically
	with a Runge-Kutta scheme as it is implemented in the
	differential equation solver \emph{ode45}\cite{ode45} in Octave.\cite{octave}
	The density operator $\hat \rho$ is represented on a numerical grid
	for the nuclear coordinate $q$ with 96 grid points, for each of the included states
	$\ke{n,M}$.
	The density matrix is propagated for 500\,fs,
	a duration selected to reflect the relevant timescale of the MgH$^+$ dissociation.	
	Due to the increased computational cost associated with the Lindblad equation,
	a strategy was developed where Lindblad operators could be summed ahead of time evolution,
	reducing both memory requirements and computational cost.
	See derivation and details about this strategy in the Appendix,
	section \ref{sec:reducing-lindblad-equation}.
	
	The accuracy of the method is tested against time evolution with our in-house software package QDng
	using the Chebyshev propagation method,\cite{ezer84jcp}
	and good agreement between the two was found.
	See the appendix section \ref{sec:benchmark} for benchmarking results,
	and further details about the implementation of the numerical method.

	\section{Results and Discussion}
	\label{sec:results-discussion}
	
	Initially, the system is vertically excited in full to the state $\ke{0,C}$,
	which corresponds to a dissociative state of the MgH$^+$ molecule.
	Since this state has no field excitations,
	the initial state will not itself decay to any other state.
	The vacuum electric field strength $\mathcal{E}_c$---which 
	scales the light-matter interaction strength
	according to Eq.~\eqref{eq:hamiltonian-interaction}---
	is sampled for a range of values: ${\mathcal{E}_c \in [0,6]}$\,GV/m.
	After 500\,fs of time evolution,
	the remaining population in the system 
	(the trace of $\hat \rho$) is recorded.
	This is the portion that has not been removed by absorbing potentials,
	and it measures the stability of the MgH$^+$ molecule
	over the range of light-matter interaction strengths.
	The mean lifetime of the field excitation, $\tau=1/\kappa$,
	is then varied over eight orders of magnitude.
	
\begin{figure*}
		\centering
		\includegraphics[width=14cm]{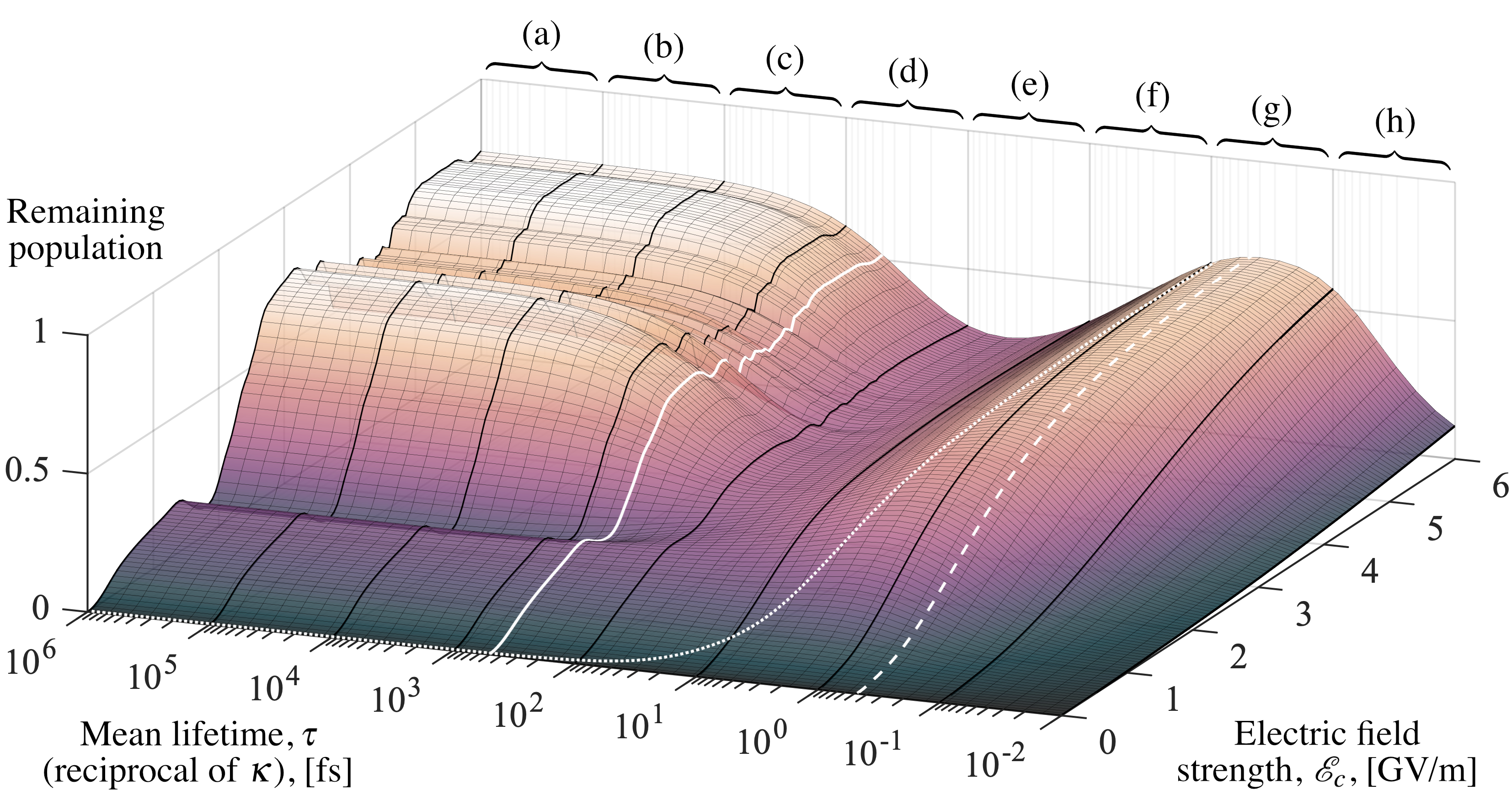}
		\caption{
		The remaining population on the vertical axis
		is the trace of $\hat \rho$,
		500\,fs of time evolution after the initial excitation to a dissociative state,
		and it measures the stability of the MgH$^+$ molecule.
		The electric field strength, $\mathcal{E}_c$, determines the strength of the light-matter coupling 
		in the interaction Hamiltonian~\eqref{eq:hamiltonian-interaction}.
		Mean lifetime, $\tau$, refers to that of the field excitation from the cavity structure,
		and $\tau = 1/\kappa$ where $\kappa$ is the decay rate
		from the Lindblad Eq.~\eqref{eq:lindblad}.
		Thicker black lines partition the lifetime parameter into eight sectors,
		(a) to (h), each corresponding to one order of magnitude.
		For reference, the solid white line in sector (d) marks the duration for time evolution (500\,fs).
		The white dashed line in sector (g) marks the time duration for light to cross the length 
		of a Fabry-Pérot type cavity (0.5\,fs).
		The dotted white line crossing several sectors marks the points where the Rabi splitting is 
		roughly equal to the full width at half maximum broadening of the field excitation,
		$\Omega_R(q) \approx \Gamma$. 
		On the left-hand side of this line resides the strong coupling regime,
		where $\Omega_R(q) \gg \Gamma$.
		\label{fig:population}}
	\end{figure*}

	Fig.\ \ref{fig:population} shows the population data from
	a batch of calculations,
	plotted on a two-dimensional grid.
	The photon lifetime $\tau$ is divided into eight sectors,
	one for each order of magnitude,
	from $10^5$\,fs in sector~(a) to $10^{-2}$\,fs in sector~(h).
	For comparison, two relevant times are marked 
	with one solid and one dashed white line.
	The solid line identifies the total duration of time evolution,
	which corresponds to 500\,fs.
	The dashed line, at 0.5\,fs, identifies the time it takes light to 
	travel across a Fabry-Pérot type cavity with length $\lambda / 2 \approx 140$\,nm.
	For such cavities, shorter lifetimes are not physical.
	
	The features calling for explanations in Fig.\ \ref{fig:population}
	occurs exclusively at higher electric field strengths.
	For small values on the other hand,
	$\mathcal{E}_c \lesssim 1$\,GV/m,
	the initial molecular excitation into the dissociative,
	but non-decaying, state $\ke{0,C}$ 
	does not exchange enough population with other dipole-coupled states,
	and essentially all the population is absorbed during the first 50\,fs.
	This is the expected behavior of a free,
	dissociating, molecule.
	The following discussion will therefore only consider 
	the behavior of a molecule coupled to the field mode,
	i.e.\ $\mathcal{E}_c \gtrsim 2$\,GV/m.
	
	\textbf{Sector (a) and (b)} 
	in Fig.\ \ref{fig:population},
	where lifetimes are long 
	($\tau$ on the order of $10^5$ or $10^4$\,fs),
	the molecule is significantly stabilized. 
	This parameter regime can be safely considered as strong coupling.
	The mean lifetime here is long enough for decay processes to be negligible,
	and the system behaves as if the lifetime was infinite (or $\kappa = 0$).
	In these sectors, a sharp rise in molecular stability can be observed 
	as the electric field strengths go beyond 2\,GV/m.
	The stability then plateaus
	and oscillates around approximately $0.7$.
	The cause for the rise in stability 
	can be attributed to the growing Rabi-splitting
	due to the stronger light-matter coupling.\cite{galego-etal-2015,kowalewski-etal-2016-jcp}
	This increases the energy difference between the polaritonic states,
	which suppresses the transfer of population between them,
	and stabilizes the molecule.
	The observed small-scale oscillations
	are understood as a consequence of 
	interference effects between nuclear wave packets in the crossing regions.\cite{wang-2007}
	The low rate of decay in these sectors retains the population in the double excitation subspace and the state does not decohere.
	Time evolution data from sector (b) is shown in Fig.\ \ref{fig:pop-b} in appendix \ref{sec:time-evolution-data}.
	
	\textbf{In sector (c)}, 
	where the mean lifetime is measured in thousands of femtoseconds
	($\tau$ on the order of $10^3$\,fs)
	the impact of photon decay starts to become noticeable.
	However, when compared to the infinite lifetime case,
	the sector is still qualitatively similar to a varying $\mathcal{E}_c$.
	The error introduced by an infinite lifetime approximation, in sector (c),
	can be quantified by population deviations from the case of $\tau = \infty$.
	To get the worst-case error in the sector for the population,
	the lifetime is fixed at its shortest, $\tau = 1000$\,fs,
	and the deviation from $\tau = \infty$ is calculated for all such points,
	which gives in an average deviation of approximately $0.07.$
	
	The analysis is repeated for a lifetime one order of magnitude larger than the timescale of the studied phenomena.
	Here 500\,fs of time evolution gives $\tau = 5000$\,fs.
	The average deviation is then calculated to approximately 0.02,
	which can be an acceptable deviation for many applications.
	
	\textbf{In sector (d)},
	with mean lifetimes on the order of hundreds of femtoseconds
	($\tau$ on the order of $100$\,fs),
	the nuclear dynamics is now on the same time scale as
	the photon decay and can not be neglected anymore.
	A small change in the lifetime now changes the stability of the molecule.
	In previous sectors, (a) to (c),
	the long lifetime implies that dissociation is primarily happening 
	in the double excitation subspace,
	but this changes in sector (d).
	The decay rate is now sufficient for the system to 
	also dissociate from	 the single excitation and ground-state subspaces
	via the states $\ke{0,A}$, $\ke{0,B}$, $\ke{1,X}$, and $\ke{0,X}$.
	For $\mathcal{E}_c \gtrsim 2.5$\,GV/m
	this type of dissociation dominates in sector (d).
	The loss of coherence in the state $\hat\rho(t)$ results in
	a suppression of the interference effects
	and the finer oscillations are dampened.
	The trapping in polariton states, which has contributed
	to the stabilization now becomes less efficient. 
	The photon decay projects the wave packet partially onto unbound vibrational eigenstates
	in the single excitation subspace,
	as well as the ground state.
	The population remaining in these unbound states for long enough will then contribute to dissociation.
	
	\textbf{In sector (e)},
	where the lifetime is on the order of tens of femtoseconds
	($\tau$ on the order of $10^1$\,fs),
	the local minimum in Fig.\ \ref{fig:population} can be understood as
	an optimum for dissociation via both the single excitation subspace and the ground state subspace.
	All subspaces are populated just long enough
	such that each subspace can contribute to the dissociation.
	In this sector, the behavior of the system 
	is dominated by effects from decay,
	and we are now firmly in the dissipative regime.
	See Fig.\ \ref{fig:pop-e}, in the appendix \ref{sec:time-evolution-data},
	for population data from time evolution in the product states.
	
	\textbf{In sector (f) and (g)},
	where the photon lifetime
	is on the order of femtoseconds or tenths of femtoseconds 
	($\tau$ on the order of $10^0$ or $10^{-1}$\,fs)
	the lifetime is short enough
	that the population decays into the ground state
	before there is enough time to contribute to dissociation from the other subspaces.
	The time evolution is shown in Fig.\ \ref{fig:pop-g}
	in appendix \ref{sec:time-evolution-data},
	where the initial population in $\ke{0,C}$ decays
	almost instantaneously to the mostly bound vibrational states in $\ke{0,X}$ 
	(via $\ke{0,A}$).
	
\begin{figure}
		\centering
		\includegraphics[width=\columnwidth]{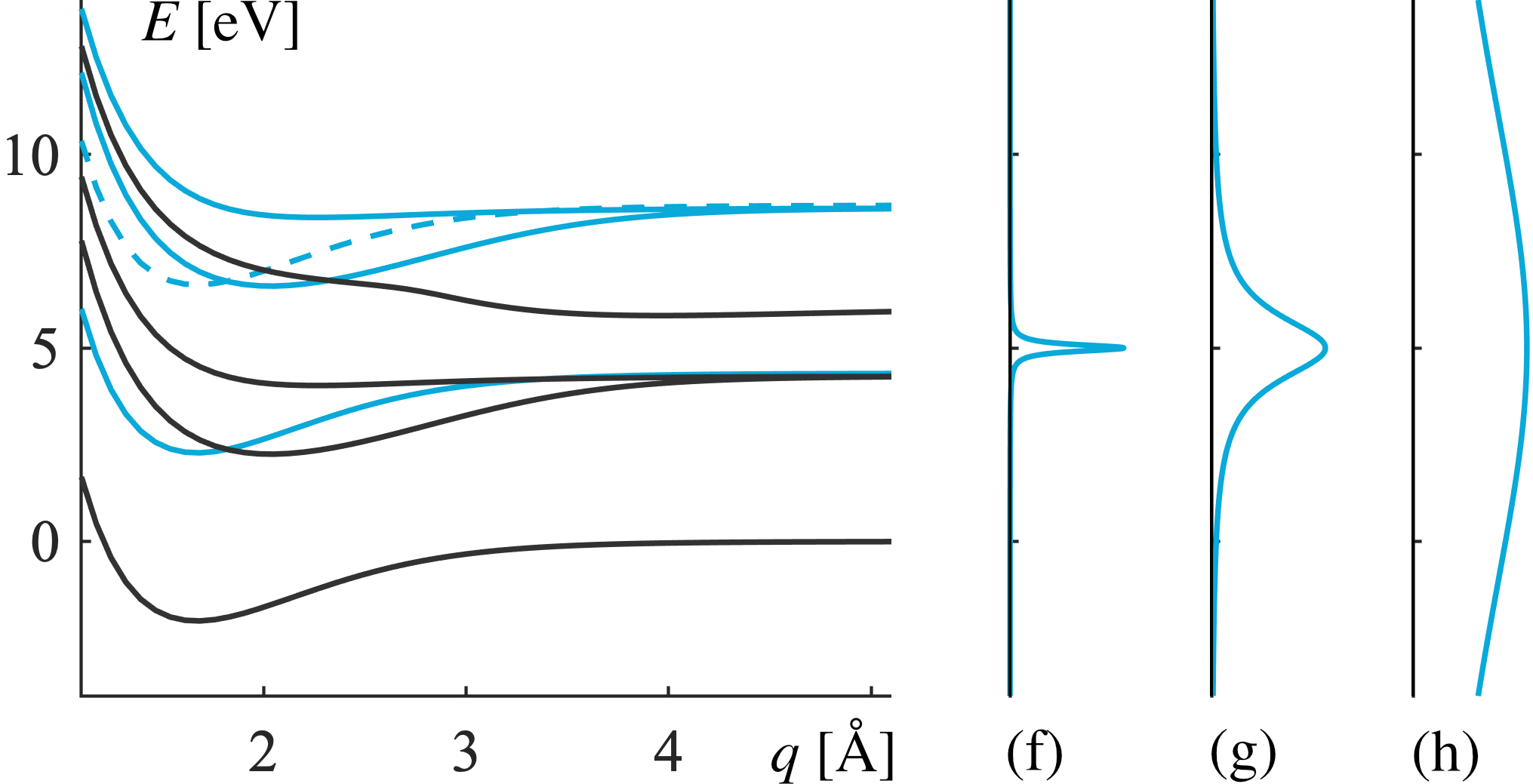}
		\caption{The left-hand side of the figure shows the basis-states 
		for the system (the same as Fig.\ \ref{fig:basis}).
		States in blue have some of its energy in the field mode.
		With the finite photon lifetime,
		their energy will spread over a superposition of energies,
		in a Lorentzian broadening.
		The degree of broadening is shown on the right-hand side,
		plotted on the same energy scale as the left-hand side
		(and for clarity renormalized to equal amplitudes).
		The graph (f) above corresponds to the lifetime $\tau = 3.2$\,fs,
		which is the logarithmic center of sector (f) in Fig.~\ref{fig:population},
		thus estimating the typical energy broadening that sector.
		The pattern continues, 
		with the graph (g) estimating the broadening in sector (g),
		and the graph (h) estimating the broadening in sector (h).
		The dashed state above has two photons,
		and thus twice the broadening of solid blue states.
		\label{fig:boradening}}
	\end{figure}

	\textbf{In sectors (g) and (h)},
	where the lifetime of field excitations 
	is on the order of tenths or hundredths of femtoseconds
	($\tau$ on the order of $10^{-1}$ or $10^{-2}$\,fs),
	a new phenomenon has to be introduced to understand why 
	the molecular stability is declining for the fastest decay.
	Basis states with a non-zero photon number
	inherit the relevant lifetime,
	and here it is short enough for the energy of these states to become a non-negligible superposition of energies,
	according to a Lorentz distribution with the full width at half maximum $\Gamma = \hbar \kappa$.
	Or put differently, states with short lifetimes experience energy broadening.
	Which states are affected,
	and the extent of this broadening is shown in Fig.~\ref{fig:boradening}.
	The superposition of energies means that states 
	which were otherwise resonantly dipole-coupled 
	acquire an increasing average detuning,
	and population transfer is suppressed.
	The outcome is a smaller spectral overlap
	and a diminished population transfer between the (sharp) initial state $\ke{0,C}$
	and the (broadened) decaying states.
	This keeps most of the population on the potential energy surface of $\ke{0,C}$,
	where it is quickly absorbed at the end of the grid,
	realizing the declining stability of the MgH$^+$ molecule
	that is most pronounced in sector (h) of Fig.~\ref{fig:population}.
	
	The increase in energy broadening from in Fig.\ \ref{fig:boradening} implies that 
	additional states will overlap in energy
	and thus their interactions should be included in the model,
	which occurs about half-way through sector (g) and certainly in sector (h).
	This means that the initial assumption about independent subspaces (see Fig.\ \ref{fig:basis})
	will start to break down.
	Fortunately, the decline in population
	seen in sector (h) from Fig.\ \ref{fig:population}
	is explained by the weaker coupling between the non-broadening initial state $\ke{0,C}$
	and the decaying states.
	This suppression will be unaffected by any missing couplings,
	and therefore, the qualitative behavior
	with a decline in stability for very short lifetimes will still~hold.
	
	Changes in chemical properties due to the interaction with lossless optical cavities 
	have been under intense theoretical study in the last decade.\cite{yuen-zhou-etal-2018,mandal-huo-2020}
	However, from this study it is clear 
	that long lifetimes are not a prerequisite for interesting modifications of chemical properties.
	In Fig.\ \ref{fig:population} a dotted white line is drawn where 
	the energy-width of the field excitation $\Gamma = \hbar \kappa$,
	is roughly equal to the Rabi splitting $\Omega_R(q) = 2\mathcal{E}_c\mu(q)$.
	Transition dipole moments are shown in Fig.~\ref{fig:trdm} as functions of $q$.
	For finding $\Omega_R(q) \approx \Gamma$ 
	the transition dipole moment is thus estimated as $\mu = 1\times10^{-29}$\,Cm,
	and a dotted white line for $\Omega_R = \Gamma$ can be drawn.
	Some orders of magnitude to the left of this line lie the strong coupling regime,
	where $\Omega_R(q) \gg \Gamma$,
	and to its right,
	energy broadening is the dominant effect.
	Still, an equally significant modification of the MgH$^+$ stability is observed in this regime.
	
	Using the same estimate for the transition dipole moment ($\mu = 1\times10^{-29}$\,Cm),
	the entire parameter regime shown in Fig.\ \ref{fig:population} falls just below the Ultrastrong Coupling regime (where $\E_c \mu /\omega_c > 0.1$),
	thus motivating the approximations made in the Hamiltonian (see section \ref{sec:hamiltonian}).

	\section{Conclusion}
	\label{sec:conclusion}
	We have studied the photostability of MgH$^+$ in a lossy cavity. 
	The dynamics have been simulated by performing nuclear wave packet dynamics via the Lindblad equation.
	This approach includes the vibronic decoherence caused indirectly by the loss of photons from the cavity.
	The studied parameter range includes decay rates inside and outside the strong coupling regime.
	
	Deep in the strong coupling regime, for cavity lifetimes longer than the nuclear dynamics ($\tau \gg 100$\,fs),
	a stabilization of MgH$^+$ is achieved through the formation of well-separated polariton states. Interference effects at the curve crossing can be observed due to the nearly
	fully coherent time evolution.
	For cavity decay rates on the order of tens of femtoseconds the stabilization effect decreases
	and the aforementioned interference effects disappear. Even though this regime can still be regarded
	as the strong coupling regime, the coherent wave packet time evolution now competes with the
	effects from dissipation. The separation of polaritonic states, which is responsible for the stabilization
	is now affected by the photon decay.  
	
	Shortening the photon lifetime even further, down below 1\,fs, increases
	the stabilization again, 
	as the molecule is rapidly funneled back into its ground state of the system.
	Our results suggest that, for an optimal photon lifetime in this region,
	low Q-factor cavities may facilitate control of photochemical reactions,
	\cite{felicetti-etal-2020}
	where several competing mechanisms are responsible for the observed phenomena:
	The population is transferred efficiently between the polariton states, which may be 
	interpreted as an optimized spectral overlap of the cavity mode \cite{Motsch2010njp} 
	with the energy width of the nuclear wave packet.
	The short photon lifetime contributes to the cooling of the system. 
	With fast enough decay,
	the molecule
	can dissipate the energy stored in the electronic excitation before
	dissociation takes place. 
	The combination of both effects results in an optimal stabilization of MgH$^+$ in this particular configuration.
		
	It is noteworthy that the stability optimum is at the border of the strong coupling regime,
	which has also been found in other studies. \cite{felicetti-etal-2020}
	This suggests that it may be an interplay of strong coupling and cavity cooling which
	is required to explain polaritonic chemistry experiments.

	\section{Acknowledgments}
	\label{sec:ack}
	This project has received funding from the European Research Council (ERC) 
	under the European Union’s Horizon 2020 research and innovation program (grant agreement No. 852286).

	\section{Data availability}
	The data that support the findings of this study are available from the corresponding author upon reasonable request.

	\section{Appendix}
	\subsection{The Hamiltonian}
	\label{sec:hamiltonian}
	The molecular Hamiltonian is here expressed in terms of 
	the four lowest electronic states,
	$\{\ke{X}$, $\ke{A}$, $\ke{B}$, $\ke{C}\}$.
	\begin{equation}
	\begin{gathered}
		\label{eq:hamiltonian-molecule}
		\hat H_m = \\[-4pt] =
		-\frac{\hbar^2}{2M}
		\frac{\d^2}{\d q^2}
		+
		V_X(q) \ke{X}\!\br{X}
		+
		V_A(q) \ke{A}\!\br{A}
		+ \\ +
		V_B(q) \ke{B}\!\br{B}
		+
		V_C(q) \ke{C}\!\br{C}
	\end{gathered}
	\end{equation}
	The reduced mass, $M$, of the MgH$^+$ molecule is 1763.2\,$m_e$.
	1D potential energy surfaces $\{V_X(q), \cdots, V_C(q)\}$
	are calculated under the Born-Oppenheimer approximation
	with the program package Molpro\cite{molpro-2006} 
	at the CASSCF(12/10)/MRCI/ROOS level of theory,
	\cite{widmark-1990,widmark-1991}
	and interpolated using splines to a grid with 96 grid-points.
	Wave-packets approaching the right-hand side of the grid are removed by 
	Gaussian-shaped absorbing potentials.
	How much norm that has been removed is stored as a function of time,
	separately for each product state 
	(see Fig.~\ref{fig:pes}).
	
	The frequency of the fundamental cavity mode $\omega_c$ 
	is chosen to match a bright transition of the Mg atom 
	(at 285\,nm).\cite{davidsson-kowalewski-2020}
	The cavity mode is modeled in the Fock basis
	$\{\ke{0}, \ke{1}, \ke{2}, \cdots\}$.
	With the photonic creation and annihilation operators, $\hat a^\dagger$ and $\hat a$ respectively,
	the cavity Hamiltonian is formed.
	\begin{equation}
		\label{eq:hamiltonian-cavity}
		\hat H_c =
		\hbar \omega_c \, \hat a^\dagger\! \hat a
	\end{equation}
	
\begin{figure}
		\centering
		\includegraphics[width=\columnwidth]{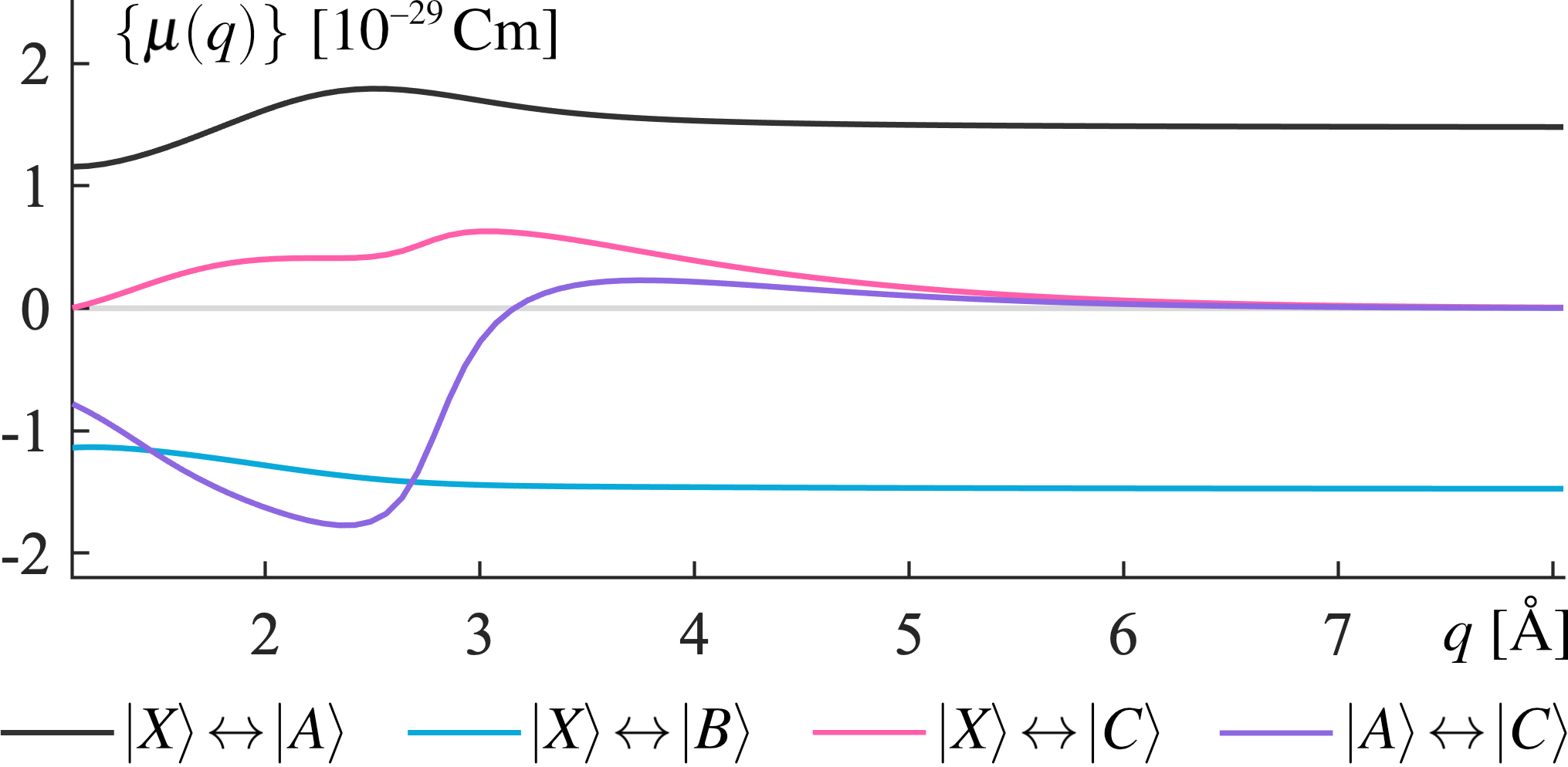}
		\caption{Four transition dipole moments are used in the calculations.
		$\ke{X} \leftrightarrow \ke{A}$, $\ke{X} \leftrightarrow \ke{B}$,
		and $\ke{A} \leftrightarrow \ke{C}$ are resonant with the cavity mode frequency
		and thus included in the interaction Hamiltonian,
		Eq.\ \eqref{eq:hamiltonian-interaction}.
		The transition $\ke{X} \leftrightarrow \ke{C}$ is used to create the initial excited state.
		\label{fig:trdm}}
	\end{figure}

	Light-matter interactions are modeled under the dipole and rotating wave approximations.
	Molecular transitions resonant with the cavity mode frequency are 
	${\ke{X} \!\leftrightarrow\! \ke{A}}$,
	${\ke{X} \!\leftrightarrow\! \ke{B}}$, and
	${\ke{A} \!\leftrightarrow\! \ke{C}}$ (see Fig.\ \ref{fig:pes} 
	for their potential energy surfaces).
	\begin{equation}
		\label{eq:hamiltonian-interaction}
		\begin{alignedat}{4}
			\hat H_{cm} = \; & \E_c \, \mu_{XA}(q)
			&&\Big(\hat a^\dagger \ke{X}\!\br{A} &&+ \hat a \, \ke{A}\!\br{X} \Big)
			&&+ \\ + \;
			&\E_c \, \mu_{XB}(q)
			&&\Big(\hat a^\dagger \ke{X}\!\br{B} &&+ \hat a \, \ke{B}\!\br{X} \Big)
			&&+ \\ + \;
			&\E_c \, \mu_{AC}(q)
			&&\Big(\hat a^\dagger \ke{A}\!\br{C} &&+ \hat a \, \ke{C}\!\br{A} \Big)
		\end{alignedat}
	\end{equation}
	The transitions $\ke{A} \!\leftrightarrow\! \ke{B}$,
	$\ke{B} \!\leftrightarrow\! \ke{C}$, 
	and $\ke{X} \!\leftrightarrow\! \ke{C}$ are not resonant with the cavity mode frequency,
	but the transition $\ke{X} \!\leftrightarrow\! \ke{C}$ is used for creating the initial excited state.
	The transition dipole moments $\{\mu(q)\}$
	are obtained at the same level of theory as the potential energy surfaces,
	and Fig.~\ref{fig:trdm} shows the ones that are used in the model.
	For a minimal model of the molecule,
	permanent dipole-moments and self-energy is omitted.
	The vacuum electric field strength,
	$\mathcal{E}_c$,
	determines the strength of the light-matter interaction
	and depends on the mode volume of the cavity structure,
	$V$.
	\begin{equation}
		\mathcal{E}_c
		=
		\sqrt{\frac{\hbar \omega_c}{2\epsilon_0 V}}
	\end{equation}

	\subsection{Summing Lindblad operators ahead of time evolution}
	\label{sec:reducing-lindblad-equation}
	
	To do time evolution with the Lindblad Eq.\ \eqref{eq:lindblad},
	all $\{\hat L_{n}\}$ operators must be stored in memory,
	and given $N$ such operators,
	$6N$ matrix multiplications are required for each time-step.
	A method for reducing memory requirements and computational cost is here demonstrated.
	Memory storage is reduced from $N$ matrices to 2.
	The computational cost goes from $6N$ matrix multiplications to
	4 matrix multiplications and 2 operations where off-diagonal elements are set to zero.
	Three assumptions are used,
	which are fulfilled in this study.
	
	\textbf{Assumption 1.} In the basis employed for the computation,
	each Lindblad operator is decaying a single basis-state to another basis-state.
	$\square$
	
	\textbf{Assumption 2.}\ There are no Lindblad pure de-phasing operators in the model.~$\square$
	
	\textbf{Assumption 3.}\ Lindblad operators and decay rates are time-independent.~$\square$
	
	If assumptions 1 and 2 do not hold this method may still be useful;
	in combination with a change of basis,
	and/or treating de-phasing operators separately.
	
	There are two useful notations for the set of all Lindblad operators.
	The most minimal notation is to simply enumerate them.
	\begin{equation}
		\big\{\hat L_n:n \in [1,\cdots,N]\big\}
	\end{equation}
	But according to assumption 1,
	each operator involves only two states from the employed basis $\{\ke{e_i}\}$.
	Thus Lindblad operators can also be expressed in terms of the basis states that comprise~them.
	\begin{equation}
		\label{eq:lindblad-operator-form}
		\big\{\hat L_{ij}:
		\hat L_{ij} \defeq \ke{e_i} \br{e_j}\big\}
	\end{equation}
	Associated decay rates are denoted with $\{\kappa_n\}$ or $\{\kappa_{ij}\}$ respectively.
	Depending on whether the emphasis is on the number of operators,
	or on the basis states that comprises them,
	these two notations will be used interchangeably in the following.
	
	The focus is on treating the Lindblad operators,
	inside the sum of the Lindblad equation.
	\begin{equation}
		\label{eq:lindblad-appendix}
		\partial_t \hat \rho
		=
		-\frac{\i}{\hbar} \big[ \hat H, \hat \rho \big]
		+
		\sum_{n}\! 
		\kappa_{n} \Big(
			\hat L_{n}\hat \rho \hat L^\dagger_{n}
			- \frac{1}{2}
			\big[ \hat L_{n}^\dagger \hat L_{n}, \hat \rho \big]_{+}
		\Big)
	\end{equation}
	These two terms in Eq.\ \eqref{eq:lindblad}
	are performing the functions of repopulating states,
	and depopulating states, in that order of appearance.
	
	The first term, responsible for repopulating states, can unfortunately not be factorized.
	\begin{equation}
		\label{eq:factorisation-not-possible}
		\sum_{n}
		\kappa_{n}
		\hat L_{n}\hat \rho \hat L^\dagger_{n}
		\,\neq\,
		\bigg(
			\sum_{n}
			\sqrt{\kappa_{n}}
			\hat L_{n}
		\bigg)
		\hat \rho
		\bigg(
			\sum_{n}
			\sqrt{\kappa_{n}}
			\hat L_{n}
		\bigg)^{\!\!\dagger}
	\end{equation}
	But we treat it as if it could be anyway,
	and define the left parenthesis from Eq.\ 
	\eqref{eq:factorisation-not-possible} as $\hat S_1$.
	\begin{equation}
		\label{eq:def-s1}
		\hat S_1
		\defeq
		\sum_{n} \sqrt{\kappa_{n}} \, \hat L_{n}
		\equiv
		\sum_{\{ij\}} \sqrt{\kappa_{ij}} \, \hat L_n
	\end{equation}
	The multiplication from Eq.\ \eqref{eq:factorisation-not-possible} is then expanded.
	\begin{equation}
		\label{eq:unwanted-cross-terms}
		\begin{alignedat}{4}
			& && \hspace{30pt}
			\hat S_1 \hat \rho \hat S_1^\dagger
			= \\
			&
			\big( \kappa_1 \hat L_1 \hat \rho \hat L_1^\dagger
			&& +
			\sqrt{\kappa_1\kappa_2} \, \hat L_1 \hat \rho \hat L_2^\dagger
			&& +
			\sqrt{\kappa_1\kappa_3} \hat L_1 \hat \rho \hat L_3^\dagger
			&& + \cdots \big) \, + \\
			&
			\big( \sqrt{\kappa_2\kappa_1} \, \hat L_2 \hat \rho \hat L_1^\dagger
			&& +
			\kappa_2 \hat L_2 \hat \rho \hat L_2^\dagger
			&& +
			\sqrt{\kappa_2\kappa_3} \, \hat L_2 \hat \rho \hat L_3^\dagger
			&& + \cdots \big) \, + \\
			&
			\big( \sqrt{\kappa_3\kappa_1} \, \hat L_3 \hat \rho \hat L_1^\dagger
			&& +
			\sqrt{\kappa_3\kappa_2} \, \hat L_3 \hat \rho \hat L_2^\dagger
			&& +
			\kappa_3 \hat L_3 \hat \rho \hat L_3^\dagger
			&& + \cdots \big) + \cdots \hspace{-30pt}
		\end{alignedat}
	\end{equation}
	Terms with a single $\kappa_n$ are identical to the desired terms 
	from the Lindblad Eq.\ \eqref{eq:lindblad-appendix},
	but there are also several unwanted cross-terms.
	Expressing  $\hat \rho$ in the employed basis $\{\ke{e_i}\}$,
	and using the result that all operators $\{\hat L_{n}\} \equiv \{\hat L_{ij}\}$ are on the form of 
	Eq.\ \eqref{eq:lindblad-operator-form},
	we simplify the operator for each type of term,
	beginning with the single~$\kappa_n$ term.
	\begin{equation}
		\label{eq:term-single-kappa}
		\hat L_{ij}\hat \rho \hat L^\dagger_{ij}
		= 
		\ke{e_i} \br{e_j}
		\bigg( \sum_{k,l} \ke{e_k} \br{e_l} \rho_{kl} \bigg)\,
		\ke{e_j} \br{e_i}
		=
		\ke{e_i} \br{e_i} \rho_{jj}
	\end{equation}
	These desired terms thus correspond to increasing population in diagonal matrix elements of $\hat \rho$
	(due to $\ke{e_i} \br{e_i}$),
	in proportion to other diagonal matrix elements of $\hat \rho$ (due to $\rho_{jj}$).
	Then the cross-terms are simplified.
	\begin{equation}
		\label{eq:term-cross}
		\hat L_{ij} \hat \rho \hat L_{mn}^\dagger
		=
		\ke{e_i} \br{e_j} 
		\bigg( \sum_{k,l} \ke{e_k} \br{e_l} \rho_{kl} \bigg)\,
		\ke{e_n} \br{e_m}
		=
		\ke{e_i} \br{e_m} \rho_{jn}
	\end{equation}
	The $\{\hat L_{ij}\}$ operators are required to be orthogonal,
	which means that either $i \neq m$ or $j \neq n$.
	I.e.\ either $\ke{e_i} \br{e_m}$ is an off-diagonal matrix element after matrix multiplication,
	or $\rho_{jn}$ is an off-diagonal matrix element before matrix multiplication (or indeed both).
	
	Setting off-diagonal elements of $\hat \rho$ to zero both before and after matrix multiplication with $\hat S_1$
	will therefore remove only the cross-terms from Eq.~\eqref{eq:unwanted-cross-terms}.
	Let $\D[\cdot]$ be this operation, 
	which sets off-diagonal matrix elements to zero in the employed basis.
	The initial sum over multiple matrix multiplications can now be rewritten using~$\hat S_1$.
	\begin{equation}
		\sum_{n}
		\kappa_{n}
		\hat L_{n}\hat \rho \hat L^\dagger_{n}
		=
		\D\Big[\hat S_1 \hat \D[\hat \rho] \hat S_1^\dagger \Big]
	\end{equation}
	
	In contrast to the repopulating term from the Lindblad Eq.\ \eqref{eq:lindblad-appendix},
	the other term, responsible for depopulation,
	can be factorized.
	\begin{equation}
	\label{eq:lindblad-terms}
	\begin{gathered}
		\sum_{n}\! 
		- \kappa_{n} \frac{1}{2}
		\big[ \hat L_{n}^\dagger \hat L_{n}, \hat \rho \big]_{+}
		= \\ =
		\bigg(\!
			- \frac{1}{2}
			\sum_{n} \kappa_{n}
			\hat L_{n}^\dagger \hat L_{n}
		\bigg)
		\hat \rho
		+
		\hat \rho
		\bigg(\!
			- \frac{1}{2}
			\sum_{n} \kappa_{n}
			\hat L_{n}^\dagger \hat L_{n}
		\bigg)
	\end{gathered}
	\end{equation}
	Defining the left parenthesis above as the operator $\hat S_2$,
	we can reduce this to two matrix multiplications.
	\begin{equation}
		\label{eq:def-s2}
		\hat S_2
		\defeq
		- \frac{1}{2}
		\sum_{n} \kappa_{n}
		\hat L_{n}^\dagger \hat L_{n}
		\equiv
		- \frac{1}{2}
		\sum_{\{ij\}} \kappa_{ij}
		\hat L_{ij}^\dagger \hat L_{ij}
	\end{equation}
		
	With the definition of $\hat S_1$ and $\hat S_2$ from Eqs.\ \eqref{eq:def-s1} and \eqref{eq:def-s2},
	they are calculated once ahead of time evolution and stored in memory,
	and the set $\{\hat L_n\}$ can be discarded.
	Together with the operation $\D[\cdot]$,
	the initial sum from Eq.\ \eqref{eq:lindblad-appendix}
	with $6N$ matrix multiplications,
	is reduced to 4 matrix multiplications 
	and 2 $\D$-operations where off-diagonal elements are set to zero.
	\begin{equation}
	\begin{gathered}
		\sum_{n}\! 
		\kappa_{n} \Big(
			\hat L_{n}\hat \rho \hat L^\dagger_{n}
			- \frac{1}{2}
			\big[ \hat L_{n}^\dagger \hat L_{n}, \hat \rho \big]_{+}
		\Big)
		= \\ =
		\D\Big[\hat S_1 \hat \D[\hat \rho] \hat S_1^\dagger \Big]
		+
		\hat S_2
		\hat \rho
		+
		\hat \rho
		\hat S_2^\dagger
	\end{gathered}
	\end{equation}

	\subsection{Numerical method and verification}
	\label{sec:benchmark}
	
\begin{figure}
		\centering
		\includegraphics[width=\columnwidth]{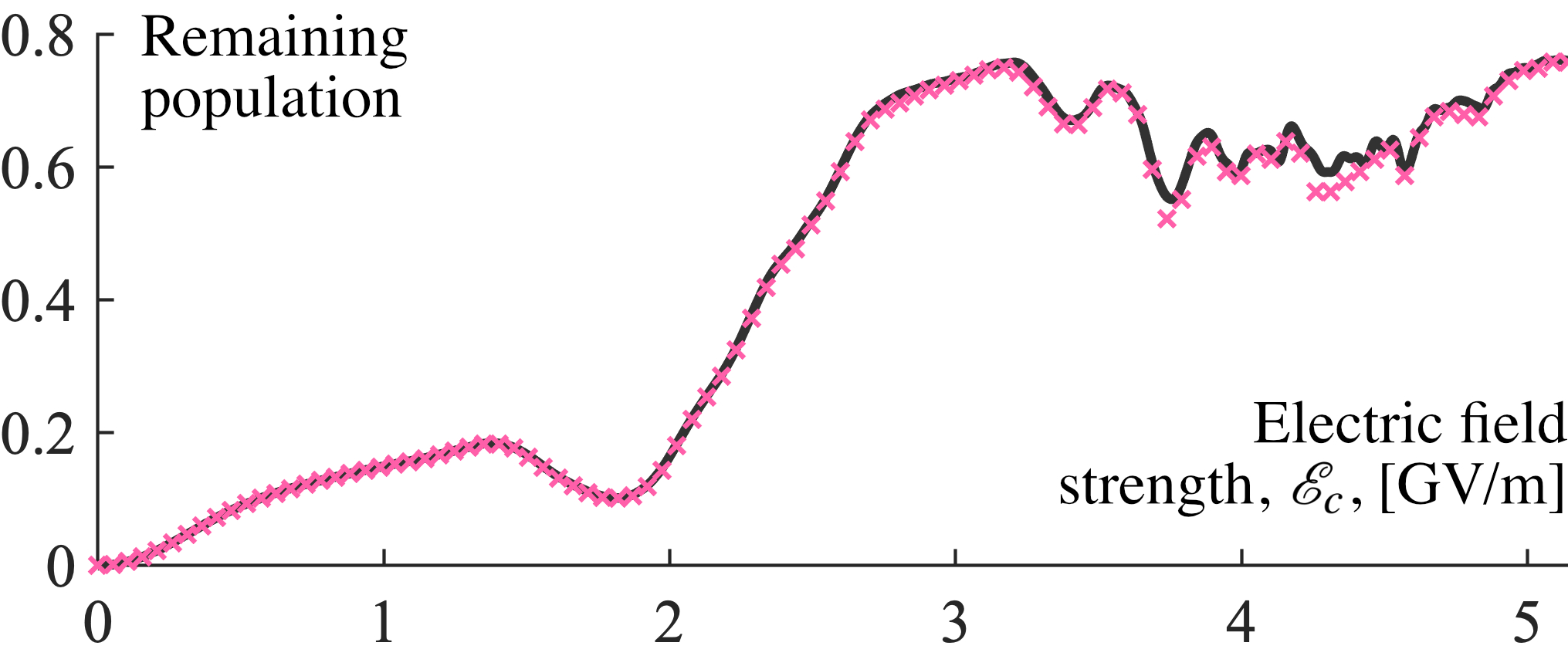}
		\caption{Data for method accuracy benchmark.
		The remaining population in the system after 500\,fs of time evolution.
		The black curve is obtained from 1000 QDng,
		Chebyshev propagator, calculations (QDng method),
		and considered the baseline for accuracy comparisons.
		Pink crosses are the results from 100 Octave, ode45, calculations (Octave method).
		\label{fig:benchmark}}
	\end{figure}

	The Lindblad equation is solved numerically,
	using the ode45\cite{ode45} (Runge-Kutta) differential equation solver,
	as implemented in Octave.\cite{octave}
	Hereafter referred to as the Octave method.	
	Verification and accuracy benchmarks of the Octave method
	are done by comparison to 
	time evolution with the Schrödinger equation,
	in the in-house QDng package,
	using the Chebyshev propagator.\cite{ezer84jcp}
	Hereafter referred to as the QDng method.
	This QDng method uses 512 grid-points 
	and a fixed duration time-step set at 0.0242\,fs, 
	previously determined to have high accuracy in similar systems
	when compared to even shorter time-steps.\cite{davidsson-kowalewski-2020}
	The Octave method uses 96 grid-points, and a variable duration time-step.
	Accuracy is here specified with a $1\times10^{-6}$ absolute tolerance,
	and $1\times10^{-3}$ relative tolerance.
	For both, methods we calculate the remaining population 
	after 500\,fs of time evolution
	(i.e.\ the same data that makes up our main result in Fig.\ \ref{fig:population}).
	The QDng method requires there to be no photon decay,
	thus $\kappa = 0$ also in the Octave method.
	A comparison of the two methods is shown in Fig.\ \ref{fig:benchmark}.
	
	At a standard deviation in the remaining population of 0.0086,
	the agreement between the two methods is satisfactory.
	Noticeable deviations occur for higher field strengths,
	where the impact of interference effects causes some disagreement.
	However, the overall profile of the oscillations is still accurately tracked.

	\subsection{Time-resolved data from simulations}
	\label{sec:time-evolution-data}
	
	Detailed time evolution data is here shown for three points in the parameter range.
	The first point is in the polaritonic strong coupling regime,
	with parameters $\mathcal{E}_c = 3.0$\,GV/m and $\tau = 3.6\times10^4$\,fs,
	directly after the sudden rise in sector (b) of Fig.~\ref{fig:population}.
	The second point has parameters $\mathcal{E}_c = 3.0$\,GV/m and $\tau = 48$\,fs,
	which puts it in the low stability region of sector (e).
	The third point has parameters $\mathcal{E}_c = 3.0$\,GV/m and $\tau = 0.58$\,fs,
	which is a point on top of the high stability region in sector (g), of Fig.~\ref{fig:population}.
	At each point, populations are plotted individually for each product state.
	See Fig.\ \ref{fig:basis} for the states.
	States which are not noticeably populated (peak population less than 0.08) are not shown.
	The total population remaining in the system is also shown,
	along with a renormalized purity.
	The renormalization is done to eliminate loss of purity caused directly from the loss of norm.
	The full time evolution is 500\,fs,
	but plots are limited to the first 250\,fs where the influential processes are taking place.
	
	\begin{figure}
		\centering
		\includegraphics[width=\columnwidth]{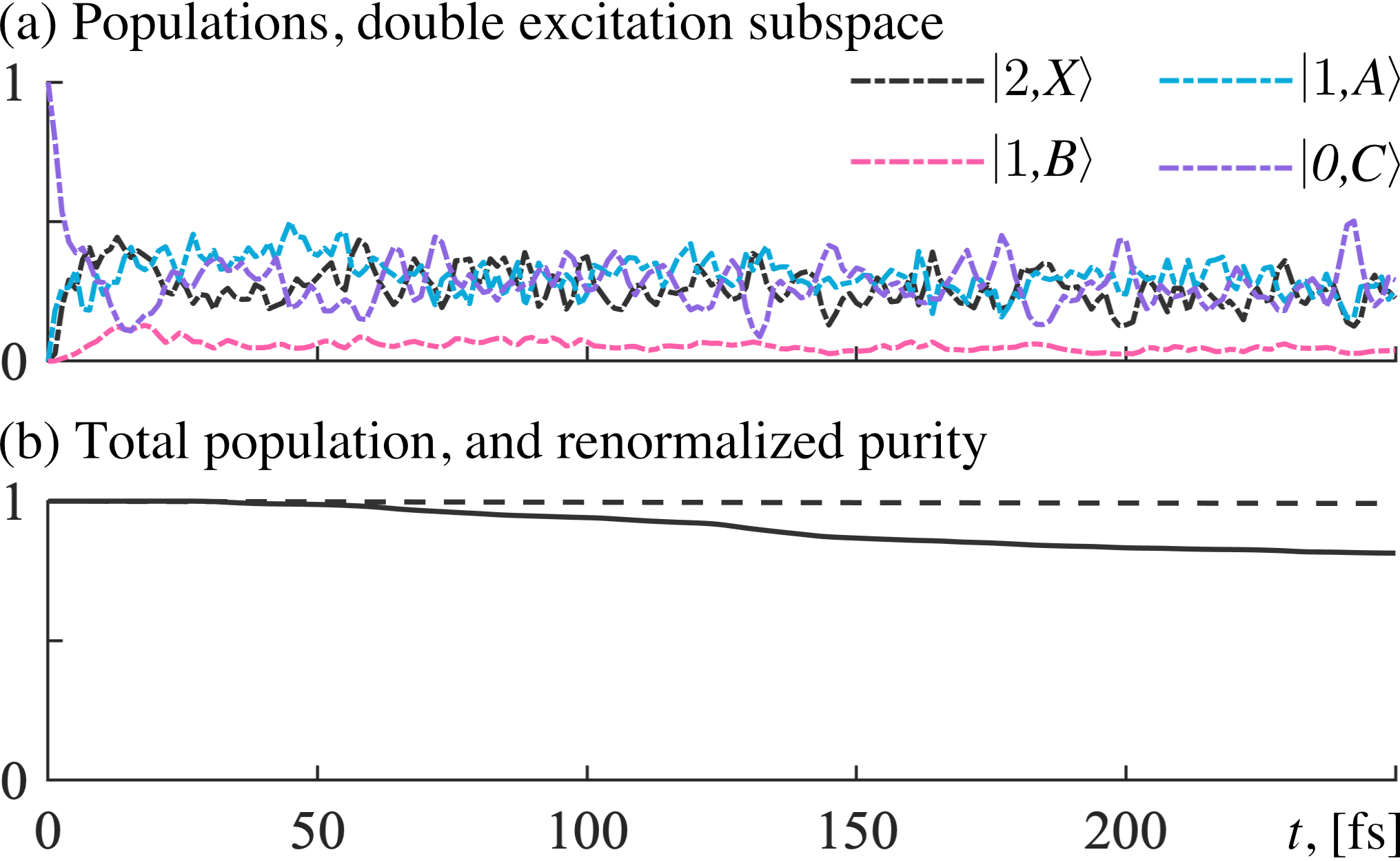}
		\caption{\label{fig:pop-b}
		Time evolution data for $\mathcal{E}_c = 3.0$\,GV/m and $\tau = 3.6\times10^4$\,fs,
		sector (b) in Fig.~\ref{fig:population}.
		States not shown has a peak population of less than 0.002.
		(a) Populated states in the double excitation subspace.
		(b) Solid line shows the total population,
		dashed line shows the renormalized purity.
		}
	\end{figure}
	
	\begin{figure}
		\centering
		\includegraphics[width=\columnwidth]{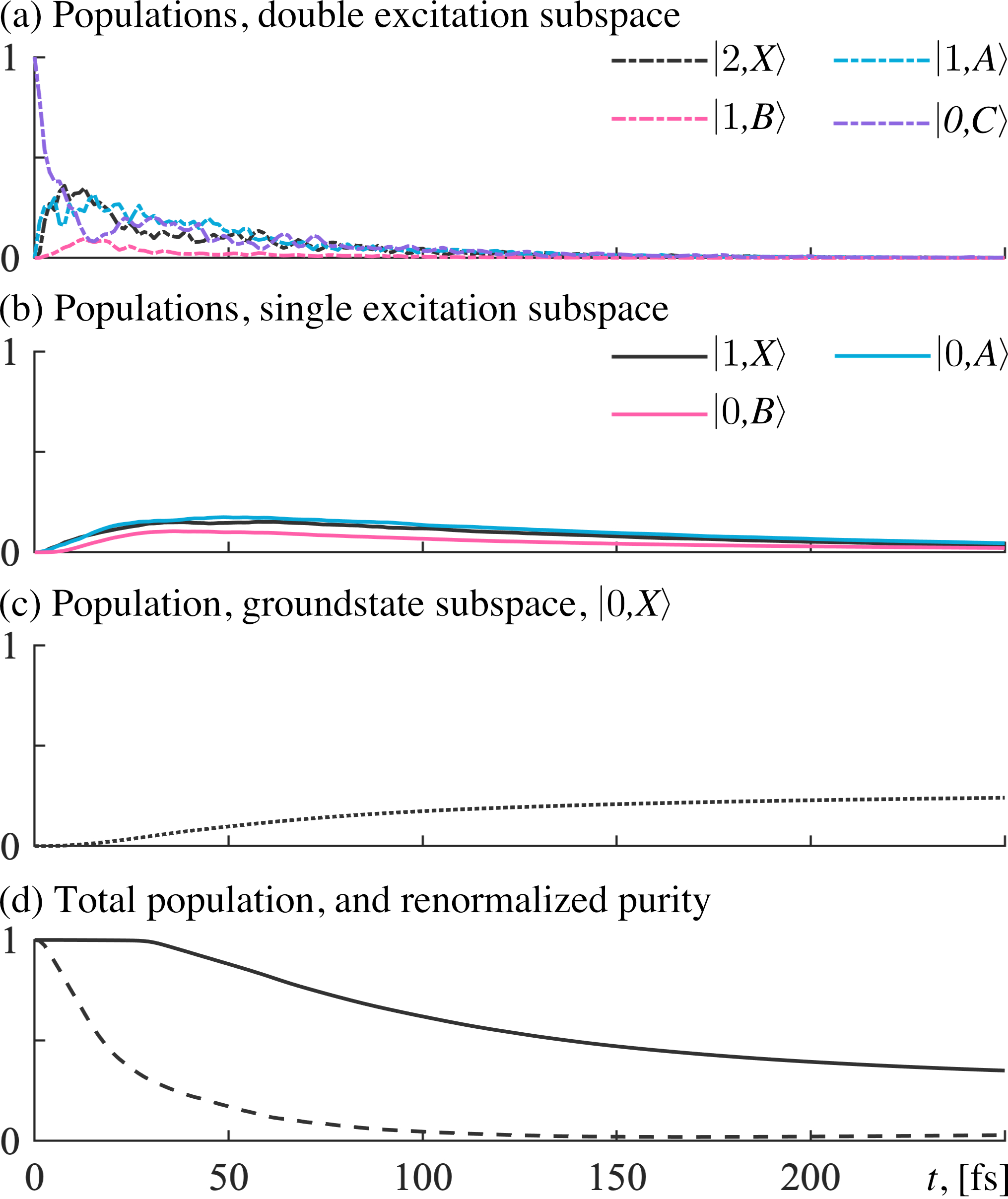}
		\caption{\label{fig:pop-e}
		Time evolution data for $\mathcal{E}_c = 3.0$\,GV/m and $\tau = 48$\,f,
		sector (e) in Fig.~\ref{fig:population}. 
		All states are noticeably populated.
		(a) Populated states in the double excitation subspace.
		(b) Populated states in the single excitation subspace.
		(c) Populated states in the ground-state subspace.
		(d) Solid line shows the total population,
		dashed line shows the renormalized purity.
		}
	\end{figure}
	
	\begin{figure}
		\centering
		\includegraphics[width=\columnwidth]{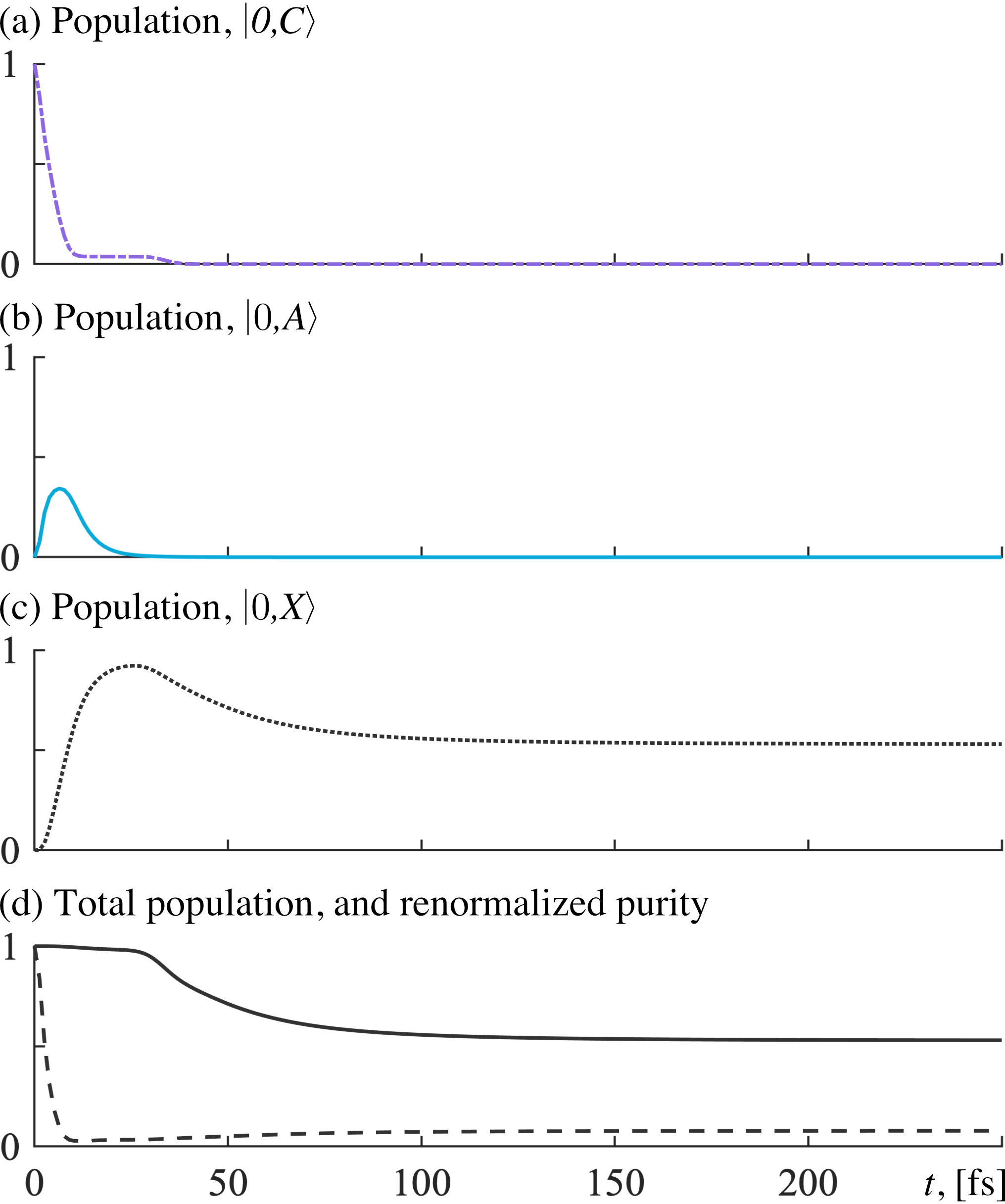}
		\caption{\label{fig:pop-g}
		Time evolution data for $\mathcal{E}_c = 3.0$\,GV/m and $\tau = 0.58$\,fs,
		sector (g) in Fig.~\ref{fig:population}.
		States not shown has a peak population of less than 0.08.
		(a) Populated states in the double excitation subspace.
		(b) Populated states in the single excitation subspace.
		(c) Populated states in the ground-state subspace.
		(d) Solid line shows the total population,
		dashed line shows the renormalized purity.
		}
	\end{figure}

	\clearpage
	\bibliography{Bibliography}

\end{document}